\documentclass[aps,prd,reprint,groupedaddress]{revtex4-1}

\usepackage[utf8]{inputenc}
\usepackage[english]{babel}
\usepackage[T1]{fontenc}
\usepackage{amsmath}
\usepackage{amsfonts}
\usepackage{hyperref}
\usepackage{amssymb}
\usepackage{graphicx}
\usepackage{tikz-feynman}
\usepackage{subfig}
\usepackage{physics}
\usepackage{array}

\begin{document}


\title{The two-body potential of Vainshtein screened theories}


\author{Adrien Kuntz}
\email[]{kuntz@cpt.univ-mrs.fr}
\affiliation{Aix Marseille Univ, Universit\'{e} de Toulon, CNRS, CPT, Marseille, France}


\date{\today}

\begin{abstract}
Adding a light scalar degree of freedom to General Relativity often induces a fifth force whose magnitude is strongly constrained by laboratory experiments and solar system tests. The Vainshtein screening mechanism ensures that the effects of this supplementary force are suppressed in dense environments. However, the field solution of theories exhibiting Vainshtein screening is only known in spherically symmetric situations. In this article we examine in different configurations the two-body potential energy of pointlike particles in a specific $P(X)$ theory with Vainshtein screening. We use ideas borrowed from the Effective One-Body approach of Buonanno and Damour in order to restrict the form of the solution. Our results indicate that, even if Vainshtein screening is also fully active in the equal-mass case, the nonlinear dependance of the two-body energy on the mass ratio implies a violation of the Equivalence Principle. We compute the contribution of this effect to the Moon orbit for generic theories equipped with Vainshtein screening.

\end{abstract}

\pacs{}

\maketitle

\section{Introduction}

The possibility of adding a new degree of freedom to General Relativity is often motivated by the cosmic acceleration. Horndeski \cite{horndeski_second-order_1974} and its recent extensions GLPV and DHOST theories \cite{gleyzes_new_2015, langlois_degenerate_2016} are prime examples of such alternative explanations to the cosmological constant. However, these theories rely on a \textit{screening mechanism} in order to hide the fifth force effects of the scalar field on solar system scales. Among these mechanisms, Vainshtein or K-mouflage screening \cite{vainshtein_problem_1972, babichev_k-mouflage_2009, Babichev:2013usa} is due to the fact that classical nonlinearities are dominant on scales smaller than the nonlinear Vainshtein radius (the quantum dominated regime, where loop corrections should be taken into account, happens at a much smaller scale \cite{deRham:2014wfa}).


In this nonlinear regime, the exact field generated by massive bodies is known only in the spherically symmetric case. Generically, it is found that the field goes as $\phi \sim r^{n}$ with $n > -1$ so that it is subdominant compared to the Newtonian gravitational potential $\Phi \sim r^{-1}$. However, tests of GR on solar system scales are very precise and provide useful bounds on these theories using the anomalous perihelion precession that they predict even if the field is screened. Moreover, the recent detection of a binary neutron star merger GW170817 both in the gravitational and electromagnetic channels \cite{the_ligo_scientific_collaboration_gw170817:_2017} has constrained the speed of gravitational waves to be that of light, killing a large set of modified gravity models \cite{creminelli_dark_2017, Ezquiaga:2017ekz, Baker:2017hug} and constraining the Vainshtein mechanism \cite{Sakstein:2017xjx, Crisostomi:2017lbg, Langlois:2017dyl}.

Focussing on the solar system for now, the simplest approximation that one can do is to treat planets as test-masses in the (screened) field generated by the sun. This implies that there is no observable effect on the precession of the perihelion of Mercury for the quartic galileon but that the cubic Galileon is already constrained by observations \cite{andrews_galileon_2013, dvali_accelerated_2003, Iorio:2012pv}. However, for comparable mass bodies the test-mass approximation is not good and this could affect the motion of bodies in a way that depend of their composition, thus violating the Equivalence Principle \cite{Belikov:2012xp, hiramatsu_equivalence_2013}. In this case, a analytic form of the two-body potential in the screened regime is missing; it could be particularly relevant for the Earth-moon system, where the Earth-Moon distance is measured with millimeter accuracy by Lunar Laser Ranging experiments \cite{Williams:2012nc, Battat_2009}.

Another important physical situation where the two-body potential could be of interest (at least in principle) is the case of two gravitationally bound compact objects (neutron stars or black holes) of comparable mass whose gravitational wave emission is now directly detected at the LIGO/Virgo interferometers. A simple order-of-magnitude estimate shows that the nonlinear Vainshtein radius for solar mass objects is $r_{*, \;1} \sim 10^{15}$ meters and $r_{*, \;2} \sim 10^{19}$ meters for two typical theories that incorporate Vainshtein screening, namely K-Essence \cite{PhysRevLett.85.4438, PhysRevD.63.103510} and Galileons \cite{nicolis_galileon_2009, Deffayet:2009mn} respectively. As the correction to any quantity like the Newtonian energy is suppressed by powers of $r/r_*$ where $r$ is the typical size of the system, this makes the effect of the scalar field on such a binary system suppressed by powers of respectively $10^{-10}$ and $10^{-14}$ for the two theories considered. As the gravitational wave phase is measured with an accuracy of $10^{-4}$ in detectors, this hides efficiently the effect of the scalar field from observations. However, the comparable mass case has not been studied in detail yet.

Futhermore, recent work on the radiation of binary systems in Galileon theories, both theoretical \cite{chu_retarded_2013, de_rham_galileon_2013, deRham:2012fw} and numerical \cite{dar_scalar_2018}, have shown that the radiation itself is screened with powers of $\lambda/r_*$, where $\lambda$ is the wavelength of the emitted radiation which is greater by an amount $1/v$ than the size of the system $r$ (here $v$ is the typical velocity of the bodies). The theoretical calculations, which are performed assuming a background field generated by a central mass $M=m_1+m_2$, are better suited for a small mass ratio inspiral, and it would be interesting to investigate on the case of nearly equal mass ratio. While we will only consider the simpler case of the conservative dynamics in this article, we find that it would be an interesting direction to continue this work.

Here we present a first step towards the study of comparable mass objects in Vainshtein screened theories, namely an approximation to the non-relativistic energy of two bodies. We first approximate the two-body energy outside of the screening radius where traditional calculations based on an expansion in terms of the interactions are reliable. We then try to obtain information on the energy in the screened region based on the knowledge that we gained from the outside. In order to restrict the form of the energy inside the screening region, we use ideas similar to the Effective One-Body (EOB) approach of Damour and Buonanno \cite{buonanno_effective_1999}. The EOB approach is the relativistic generalization of the well-known fact that the two-body problem in Newtonian mechanics is solved by relating it to the motion of a test particle of mass $\mu = \frac{m_1m_2}{m_1+m_2}$ in an external field generated by a body of mass $M=m_1+m_2$. The generalization to GR is that the two-body motion can be approximated by the motion of a test particle of mass $\mu$ in an external deformed Schwartzchild metric, the deformation parameter being the symmetric mass ratio $\nu = \frac{\mu}{M}$. This allows to provide an accurate description of the dynamics of binary systems in regimes that are inaccessible to other analytical approaches, such as Post-Newtonian or Gravitational Self-Force computations \cite{Damour:2012mv}. Concerning modified gravity, the EOB formalism has been generalized to scalar-tensor theories in \cite{Julie:2017ucp, Julie:2017pkb}.

We propose to apply this idea to a two-body system in a Vainshtein screened theory. Since it maps the two-body problem into a spherically symmetric one-body problem, it allows to recover the energy as a function of the known exact field solution. The difficulty is then to identify the parameters of this deformed exact solution. Outside the screening radius, a perturbative (Feynman) expansion allows us to identify the EOB parameters. However, the perturbative expansion breaks down for objects separated by less than the screening radius.


It is however easy to identify the EOB parameters using a simple numerical simulation, and the rest of this article will be devoted to this task. Numerical simulations of screened theories in the quasi-static limit have mostly focused on N-body simulations for cosmological applications \cite{Barreira:2013eea, Li:2011vk, Li:2013nua, Khoury:2009tk, 2009PhRvD..80l3003S, Schmidt:2009sg}, which is not our concern here. We thus implement our own code using the Finite element solver Fenics \cite{ans20553}. We first obtain the energy outside the screening radius in order to confirm our preliminary results. We then concentrate on the two-body energy in the fully screened situation which is relevant for astrophysical systems, and for arbitrary mass ratios. The most important result of this paper is the final mass ratio dependance of the two-body energy, and it is shown in Figure \ref{fig:b0_num}. We will finally derive the effect on the orbit of the Moon around the Earth of such a violation of the Equivalence Principle in Sec. \ref{sec:WEP}.

Let us now be more precise about the screening theory that we consider. We will focus on a typical theory that incorporate this effect, namely a specific form of K-Mouflage or $P(X)$ \cite{babichev_k-mouflage_2009}
\begin{equation}
S = \int d^4x \left[ - \frac{(\partial \phi)^2}{2} - \frac{1}{4\Lambda^4} (\partial \phi)^4 + \frac{\phi T}{M_P} \right] \; ,
\label{eq:k-mouflage}
\end{equation}
with two static point-like sources,
\begin{equation}
T = -m_1 \delta^3(\mathbf{x} - \mathbf{x}_1) - m_2 \delta^3(\mathbf{x} - \mathbf{x}_2) \; .
\end{equation} 
Note that we have concentrated on the scalar part of the action, the gravitational action being the one of standard GR. It would be easy to generalize ours results to other well-known theories that incorporate Vainshtein screening, e.g the cubic Galileon \cite{nicolis_galileon_2009}.

We stress that the chosen form for the interaction in the K-Mouflage case (with a negative sign in front of the quartic term) has a speed of sound around a cosmological background greater than one. Ref \cite{barreira_k-mouflage_2015} discusses the conditions needed in order to have a viable k-Mouflage theory. Nonetheless, we chose the theory \eqref{eq:k-mouflage} to test our method because it is one of the simplest settings in which there is Vainshtein screening and the numerical implementation with finite elements is easier since its (weak form) equation of motion involves only first derivatives.

This paper is organized as follows. In Sec. \ref{sec:outside} we compute the two-body energy in the weakly coupled regime where standard perturbative techniques apply. After resumming a class of Feynman diagrams, we recast the problem into an EOB framework in Section \ref{sec:EOB} and discuss the form of the EOB energy map in the strongly coupled regime. Section \ref{sec:numerical} then presents the numerical solution to the two-body problem in different regimes. We analyze the consequences of the violation of the Equivalence Principle that it implies on the Earth-Moon-Sun system in Sec. \ref{sec:WEP}, before concluding in Sec. \ref{sec:conclusion}. We use natural units where $\hbar = c = 1$.

\section{Two-body energy outside the screening radius} \label{sec:outside}

In this Section we will concentrate on the action \eqref{eq:k-mouflage}. We will only consider static point-sources, so we will ignore time from now on and focus only on the potential energy $E$. In order to have simpler expression in what follows and also to compare directly our results to the numerical simulation, we will introduce rescaled variables
\begin{equation} \label{eq:rescaling}
\tilde{\phi} = \frac{\phi}{\Lambda^2}, \quad \tilde{m} = \frac{m}{4\pi M_P \Lambda^2} \; ,
\end{equation}
so that the action of the system writes as
\begin{equation}
S = \Lambda^4 \int dt d^3x \left[ - \frac{1}{2} (\nabla \tilde{\phi})^2 - \frac{1}{4} (\nabla \tilde{\phi})^4 + \tilde{\phi} \tilde{T} \right] \; ,
\label{eq:k_mouflage_rescaled}
\end{equation}
where $\tilde{T} = -4\pi \tilde{m}_1 \delta^3(\mathbf{x} - \mathbf{x}_1) - 4 \pi \tilde{m}_2 \delta^3(\mathbf{x} - \mathbf{x}_2)$, and we will ignore the tilde from now on. Be careful that now the dimension of $m$ is (in natural units) $-2$ and the dimension of $\phi$ is $-1$.

\subsection{Spherically symmetric case}

Consider the action \eqref{eq:k_mouflage_rescaled} with a single point-like source $T = -4\pi M \delta^3(\mathbf{x})$. The equation of motion that follows from it reads
\begin{equation}
\partial_i \left(\partial_i \phi + (\partial \phi)^2 \partial_i \phi \right) = - T \; .
\end{equation}

Using spherical symmetry and integrating over a sphere, one can reduce it to a single ordinary differential equation
\begin{equation}
\phi' + (\phi')^3 = \frac{M}{r^2} \; .
\label{eq:cubic_eq}
\end{equation}

This can be solved exactly and writes, for a field vanishing at infinity,
\begin{equation}\label{eq:hypergeom}
\phi_M(r) = - \frac{M}{r} {}_3F_2 \left(\frac{1}{4}, \frac{1}{3}, \frac{2}{3} ; \frac{5}{4}, \frac{3}{2} ;- \frac{27M^2}{4r^4} \right) \; ,
\end{equation}
where ${}_pF_q$ is the generalized hypergeometric function.

The solution has two regimes separated by the nonlinear scale $r_*$
\begin{align}
\begin{split}
\phi_M &= - \frac{M}{r} + \frac{M^3}{5 r^5} + \dots, \quad r > r_* \; , \\
&= C + 3 \left(  M r \right)^{1/3} + \dots, \quad r<r_* \; ,
\end{split}
\label{eq:spherical_symmetry}
\end{align}
where $C \simeq -3.7 \sqrt{M}$ is a constant of integration (we chose the constant such that the field vanishes at infinity, so it cannot also vanish in zero), and the nonlinear scale is given by
\begin{equation}
r_* = \left( \frac{27}{4} \right)^{1/4} \sqrt{M} \; ,
\end{equation}
and corresponds to the radius of convergence of the two series written above.

These two regimes can be seen as expansions in $r_*/r$ and $r/r_*$ respectively (in fact, it is even possible to reformulate the initial action with additional fields in order to make the screened regime appear from the beginning, see \cite{gabadadze_classical_2012}). We have expanded up to next-to-leading order in the $r>r_*$ case because it will prove useful in the following. Choosing $\Lambda$ to be responsible for the cosmic acceleration, $\Lambda^2 = M_P H_0$ \cite{barreira_k-mouflage_2015}, one has $r_* \sim 0.1$ Pc for the sun, thus rendering the next order in the $r<r_*$ case very subdominant concerning Solar System experiments.

\subsection{Two-body problem} \label{sec:two_body}

Let's now take a two-body source $T = -4 \pi m_1 \delta^3(\mathbf{x} - \mathbf{x}_1) - 4 \pi m_2 \delta^3(\mathbf{x} - \mathbf{x}_2)$. The salient feature of the two-body problem in screened theories is that one can not compute the energy by superposing one-body energies, as one usually does in Newtonian gravity. Said differently, one can not recast the problem in terms of a simple ODE, and instead one should solve a nonlinear PDE of 2 variables, thus explaining the lack of analytical results. Nonetheless, the problem is well formulated outside the screening radius where the nonlinear term can be treated as an interaction. Let us now imagine to have two bodies outside of their respective screening radius and calculate the first nonlinear correction to the potential energy of the two objects. To this aim, we will use Feynman rules derived from the action \eqref{eq:k_mouflage_rescaled}. The potential energy of the system is $\int dt E=-S_{cl}$ where the classical action is obtained as the saddle-point of the path integral,
\begin{equation}
e^{iS_\mathrm{cl}[\mathbf{x}_1, \mathbf{x}_2]} = \int \mathcal{D}[\phi] e^{iS[\mathbf{x}_1, \mathbf{x}_2, \phi]} \; .
\end{equation}

Before we start, let us give directly the main result of this Section : the gravitational energy of the two bodies is
\begin{equation}
\frac{E}{4\pi \Lambda^4} = - \frac{m_1 m_2}{r} + \frac{m_1m_2(m_1^2+m_2^2)}{5 r^5} + \dots \; ,
\label{eq:2_body_energy}
\end{equation}
where $r = |\mathbf{x}_1 - \mathbf{x}_2|$. We have factored out a multiplicative coefficient of $\Lambda^4$ coming directly from eq. \ref{eq:k_mouflage_rescaled}: for convenience we will also use a rescaled energy defined by
\begin{equation}
\tilde{E} = \frac{E}{\Lambda^4} \; ,
\end{equation}
and forget the tilde from now on.

Let us now derive this result. Introducing the Fourier modes of the field $\phi = \int_\mathbf{k} \phi_\mathbf{k} e^{i \mathbf{k} \cdot \mathbf{r}}$ (where $\int_\mathbf{k}$ means $\int \frac{d^3k}{(2\pi)^3}$), the non-relativistic propagator of the field is given by
\begin{equation}
\left\langle T \phi_\mathbf{k}(t) \phi_\mathbf{q}(t') \right\rangle = -i (2\pi)^3 \delta^3(\mathbf{k} + \mathbf{q}) \delta(t-t') \frac{1}{k^2} \; .
\end{equation}

\begin{figure}[h]
	\centering
		\begin{tikzpicture}
			\begin{feynman}
				\vertex (i1);
				\vertex [right=of i1] (a);
				\vertex [right=of a] (f1);
				\vertex [below=of i1] (i2);
				\vertex [below=of a] (b);
				\vertex [below=of f1] (f2);
				
				\diagram*{
				i1 -- [fermion] (a) -- [fermion] (f1),
				(i2) -- [fermion] (b) -- [fermion] (f2),
				(a) -- [scalar] (b)
				};
			\end{feynman}
		
		\end{tikzpicture}
	
\caption{Feynman diagram contributing to the Newtonian potential. External sources are represented as straight lines and scalars as dotted lines}
\label{fig:Newt_pot}
\end{figure}
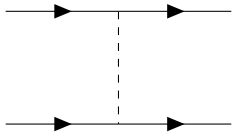

To obtain $S_{cl}$, we have to calculate all connected (once we remove the source) Feynman diagrams without internal loops, treating $(\partial \phi)^4$ and $T$ as interactions (loops would represent quantum corrections to this potential and are suppressed by powers of $\hbar/L$ where $L$ is the total angular momentum of the system \cite{goldberger_effective_2006}, so we will completely ignore them). The full Feynman rules of a simple conformally coupled scalar field in the presence of gravity are given in Ref. \cite{Kuntz:2019zef}. At lowest order, there is the one-scalar exchange of Fig \ref{fig:Newt_pot} that gives rise to the Newtonian potential between the two sources,
\begin{equation}
\frac{E_\mathrm{Newt}}{4\pi} = -\frac{m_1 m_2}{r} \; .
\label{eq:Newtonian_energy}
\end{equation}

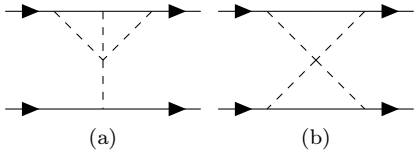
\begin{figure}
	\centering
	
	\subfloat[]{
		\begin{tikzpicture}
			\begin{feynman}
				\vertex (i1);
				\vertex [right=2em of i1] (a);
				\vertex [right=2em of a] (t1);
				\vertex [right=2em of t1] (b);
				\vertex [right=2em of b] (f1);
				\vertex [below=2em of t1] (c);
				\vertex [below=2em of c] (t2);
				\vertex [left=2em of t2] (d);
				\vertex [left=2em of d] (i2);
				\vertex [right=2em of t2] (e);
				\vertex [right=2em of e] (f2);
				
				\diagram*{
				(i1) -- [fermion] (a) -- (t1) -- (b) -- [fermion] (f1),
				(a) -- [scalar] (c),
				(b) -- [scalar] (c),
				(t1) -- [scalar] (c),
				(c) -- [scalar] (t2),
				(i2) -- [fermion] (d) -- (t2) -- (e) -- [fermion] (f2),
				};
				
			\end{feynman}
			\label{subfig:correction_kmouflage_4a}
		
		\end{tikzpicture}
	}
	\subfloat[]{
		\begin{tikzpicture}
			\begin{feynman}
				\vertex (i1);
				\vertex [right=2em of i1] (a);
				\vertex [right=2em of a] (t1);
				\vertex [right=2em of t1] (b);
				\vertex [right=2em of b] (f1);
				\vertex [below=2em of t1] (c);
				\vertex [below=2em of c] (t2);
				\vertex [left=2em of t2] (d);
				\vertex [left=2em of d] (i2);
				\vertex [right=2em of t2] (e);
				\vertex [right=2em of e] (f2);
				
				\diagram*{
				(i1) -- [fermion] (a) -- (t1) -- (b) -- [fermion] (f1),
				(a) -- [scalar] (c),
				(b) -- [scalar] (c),
				(c) -- [scalar] (d),
				(c) -- [scalar] (e),
				(i2) -- [fermion] (d) -- (t2) -- (e) -- [fermion] (f2),
				};
				
			\end{feynman}
			\label{subfig:correction_kmouflage_4b}
		
		\end{tikzpicture}
	}

\caption{Feynman diagrams contributing to the first nonlinear correction in a K-mouflage theory. The first one should be added with its symmetric counterpart.}
\label{fig:correction_kmouflage}
\end{figure}

In order to compute the first nonlinear correction to it, we have to calculate the two diagrams of Figure \ref{fig:correction_kmouflage}. The first one can be put in the following form
\begin{align}
\begin{split}
\mathrm{Fig} \ref{subfig:correction_kmouflage_4a} &= i \frac{(4\pi)^4}{3} m_1^3m_2 \int_{\mathbf{K}, \mathbf{k}_1, \mathbf{k}_2} e^{i \mathbf{K} \cdot \mathbf{r}} \frac{1}{\mathbf{k}_1^2 \mathbf{k}_2^2 \mathbf{K}^2 (\mathbf{K}+\mathbf{k}_1 + \mathbf{k}_2)^2} \\
& \times \left( \mathbf{k}_1 \cdot \mathbf{k}_2 (\mathbf{K}+\mathbf{k}_1 + \mathbf{k}_2) \cdot \mathbf{K} + 2 \; \mathrm{perm}  \right) \; ,
\end{split}
\end{align}
where $\mathbf{r} = \mathbf{x}_1 - \mathbf{x}_2$. The remaining integrals over momentum should be computed with dimensional regularization. To this aim a set of useful integrals in $d=3-2\epsilon$ are given in Appendix \ref{sec:useful_integrals}.  One can simplify the calculation with the following observation : after having integrated $k_1$ and $k_2$, one is left with an integral over $K$ which, for dimensional reasons (the first correction to the two-body potential is proportional to $r^{-5}$, see \eqref{eq:spherical_symmetry}), is of the form
\begin{equation}
\int_{\mathbf{K}} \mathbf{K}^2 e^{i \mathbf{K} \cdot \mathbf{r}} \; ,
\end{equation}
(there could also be a factor of $\epsilon$ in the power of $\mathbf{K}^2$, but it does not change the validity of the argument). Then using formula \eqref{eq:intK}, one can see that there is a pole of the Gamma function in the \textit{denominator}. Consequently, this integral will vanish in dim. reg (it will be proportional to $\epsilon$) unless there is also a pole of the Gamma function in the numerator. Keeping the only pole that appear in the numerator, and repeatedly using the formulaes given in Appendix \ref{sec:useful_integrals}, one can find that
\begin{equation}
\mathrm{Fig} \ref{subfig:correction_kmouflage_4a} = -4\pi i \frac{m_1^3m_2}{5 r^5} \; .
\end{equation}

One can use the same machinery to calculate the second diagram \ref{subfig:correction_kmouflage_4b}. However, in this case there is no pole of the Gamma function at the numerator, and consequently this diagram vanishes in dim. reg. Including the symmetric counterpart of \ref{subfig:correction_kmouflage_4a}, we finally obtain the formula for the first correction to the two-body energy given in eq. \eqref{eq:2_body_energy}.

\subsection{Resumming the test-mass diagrams} \label{sec:resum_TM}

The result of Feynman integrals concerning diagram \ref{subfig:correction_kmouflage_4a} should come as no surprise. Indeed, as we will now explain, it should come back to the energy of a point-particle mass in case of one of the masses goes to zero. In the following, we will assume without loss of generality that $m_1 < m_2$.

Consider the limit $m_1 \rightarrow 0$. Then the two-body energy should reduce to the energy of a point-particle $m_1$ in the external field generated by $m_2$, which is (the energy being obtained from the field $\phi_\mathrm{cl}$ of eq. \eqref{eq:spherical_symmetry} via $E_\mathrm{pp} = - \int d^3x \phi_\mathrm{cl} T$)
\begin{equation}
\frac{E_\mathrm{pp}}{4\pi} = - \frac{m_1 m_2}{r} + \frac{m_1m_2^3}{5 r^5} + \dots
\end{equation}

From there we see that the result of diagram \ref{subfig:correction_kmouflage_4a}, which ultimately gives the numerical prefactor in front of the first-order correction to the two-body energy, could not have been otherwise. Had the diagram of figure \ref{subfig:correction_kmouflage_4b} been nonzero, its value would not have been fixed by this observation, because it is proportional to $m_1^2 m_2^2$, which vanish (compared to the test-mass diagram proportional to $m_1$) in the test-mass limit.

\begin{figure}
	\centering
	
		\begin{tikzpicture}
			\begin{feynman}
				\vertex (i1);
				\vertex [right=2em of i1] (a);
				\vertex [right=2em of a] (t1);
				\vertex [right=2em of t1] (b);
				\vertex [right=2em of b] (f1);
				\node [below=1em of t1, blob] (c);
				\vertex [below=2em of c] (t2);
				\vertex [left=2em of t2] (d);
				\vertex [left=2em of d] (i2);
				\vertex [right=2em of t2] (e);
				\vertex [right=2em of e] (f2);
				
				\diagram*{
				(i1) -- [fermion] (a) -- (t1) -- (b) -- [fermion] (f1),
				(t1) -- [scalar] (c),
				(c) -- [scalar] (d),
				(c) -- [scalar] (e),
				(i2) -- [fermion] (d) -- (t2) -- (e) -- [fermion] (f2),
				};
				
				\draw[] (t2.north) node[above] {$\dots$};
				
			\end{feynman}
		
		\end{tikzpicture}

\caption{Test-mass diagram with a single coupling to the first particle $m_1$ and $P$ couplings to the second particle $m_2$. The number of internal vertices (not involving particles worldlines) is $N$.}
\label{fig:test_mass}
\end{figure}
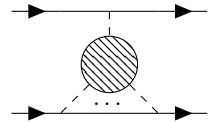

Based on this observation, we propose to resum a particular class of diagram which share the same property at any order in the nonlinear expansion, which we call test-mass diagrams. They are given by a single coupling to the mass $m_1$ and any number of couplings to $m_2$, as illustrated in Figure \ref{fig:test_mass}. Denoting by $N$ the expansion order (i.e, the number of vertex corresponding to the insertion of the nonlinear operator in the diagrams) and by $P$ the number of $m_2$ mass insertions, we have that there are $1+4N+P$ field insertions in this diagram. Since there are no loops and the diagram should be connected once we remove the particles wordlines, there are $N-1$ 'internal' propagators, i.e propagators that connect two nonlinear vertices. Finally, the total number of propagators is $P+1+(N-1)$. Since the total number of field insertions is two times the number of propagators, we get the relation
\begin{equation}
P=2N+1 \; ,
\end{equation}
so that at order $N$ the mass coefficient of this graph (and its symmetric counterpart) is $m_1m_2(m_1^{2N}+m_2^{2N})$.

One can even argue that, at a given perturbation order $N$, this test-mass diagram is the leading one away from the test-mass limit. Indeed, the contribution of other kinds of diagrams, with more $m_1$ insertions (we found the only other second-order diagram to vanish, but we see no reason why it should be the case at higher orders), would be of the form $(m_1m_2)^q(m_1^{2(N+1-q)} + m_2^{2(N+1-q)})$ where $2 \leq q \leq N+1$. For $m_2 > m_1$ and in the large $N$ limit, the ratio of this quantity to the test-mass diagram is
\begin{equation}
\left( \frac{m_1}{m_2} \right)^{q-1} \; ,
\end{equation}
which is less than one. However, this is not accounting for the fact that there can be a large number of these other diagrams, which can make them count as much as the test-mass one. Anyway, we shall content ourselves with having understood at least a part of the nonlinear energy.

Now the exact energy of a point-particle $m_1$ in the field generated by $m_2$ is
\begin{equation}
\frac{E_{pp}}{4 \pi} = m_1 \sum_{N \geq 0} \alpha_N \frac{m_2^{2N+1}}{r^{4N+1}} \; ,
\end{equation}
where $\alpha_N$ are numerical coefficients that can be easily found by solving eq. \eqref{eq:cubic_eq}. By writing in a identical way the contribution of the test-mass graphs to the two-body energy,
\begin{equation}
\frac{E}{4\pi} = \sum_{N \geq 0} \beta_N \frac{m_1m_2(m_1^{2N}+m_2^{2N})}{r^{4N+1}} \; ,
\label{eq:2_body_energy_resummed}
\end{equation}
we see that in order to have the good test-mass limit one should impose $\beta_N = \alpha_N$ for $N \geq 1$, and $\beta_0 = \frac{\alpha_0}{2} = - \frac{1}{2}$.
Denoting by $\phi_m$ the (exact) spherically symmetric field generated by a body of mass $m$ in eq. \eqref{eq:spherical_symmetry}, one finally finds for the contribution of test-mass graphs to the two-body energy
\begin{equation}
\frac{E}{4\pi} = \frac{m_1 m_2}{r} + m_1 \phi_{m_2}(r) + m_2 \phi_{m_1}(r) \; .
\label{eq:energy_outside_exact}
\end{equation}

This result can be intuitively understood as being the symmetric sum of one-body energies, plus a compensating term that ensures that the Newtonian limit $r \rightarrow \infty$ (where $\phi_m \sim -m/r$) is correct. We will compare this analytical resummed energy to the numerical solution in Sec. \ref{sec:results}.

\section{Effective One-Body approach}
\label{sec:EOB}

\subsection{Energy map outside}

Very much like in GR, where the motion of a two-body system (expanded in powers of $r_s/r$, where $r_s$ is the Schwartzchild radius of the combined mass $M=m_1+m_2$) can be recast in the motion of a test-mass in a modified external Schwartzchild metric \cite{buonanno_effective_1999}, the above formula for the two-body energy can be expressed into the energy of such a point-particle in a modified external field. Of course, what will now play the role of the Schwartzchild radius is the nonlinear radius $r_*$. Let us define, in top of the reduced mass $\mu$ and the total mass $M$ which are the two masses naturally associated to the effective problem, the mass ratio $x$ as
\begin{align}
\begin{split}
\mu &= \frac{m_1m_2}{m_1+m_2} \; , \\
M &= m_1+m_2 \; , \\
x &= \frac{m_1}{m_1+m_2} \; .
\end{split}
\end{align}

The kinetic energy of the two objects is easy to rewrite in terms of an effective kinetic energy since we have the well-known relation
\begin{equation}
\frac{1}{2} m_1 v_1^2 + \frac{1}{2}m_2 v_2^2 = \frac{1}{2} \mu v^2 \; ,
\end{equation}
where $\mathbf{v} = \mathbf{v}_1 - \mathbf{v}_2$, and we have set the center-of-mass to the origin of coordinates (this center-of-mass definition would be modified by relativistic corrections, but we do not consider these in this treatment).

As for the approximate potential energy that we obtained above \eqref{eq:2_body_energy_resummed}, it can be rewritten in terms of the effective parameters as
\begin{equation}
\frac{E}{4\pi} = \mu \sum_{N \geq 0} \beta_N \frac{M^{2N+1} \left(x^{2N}+(1-x)^{2N}\right)}{r^{4N+1}} \; .
\end{equation}

We now see that, outside the nonlinear radius, the motion of a two-body system separated by $r = |\mathbf{x}_1 - \mathbf{x}_2|$ can be identified with the motion of a test-particle of mass $\mu$ (at a distance $r$ from the origin) in a modified external field created by $M$, whose modified coefficients in the nonlinear expansion are given by
\begin{align}
\begin{split}
\tilde{\alpha}_N &= \alpha_N \left(x^{2N}+(1-x)^{2N}\right), \; N>0 \;, \\
\tilde{\alpha}_0 &= \alpha_0 \; .
\end{split}
\label{eq:modified_coeffs}
\end{align}

What is the nonlinear radius associated to the effective problem ? The perturbative expansion \eqref{eq:modified_coeffs} that we wrote above breaks down at the nonlinear radius associated to the biggest of the two masses, i.e $r_* = M(1-x)$. However this result is really tied to the test-mass resummation, and calculating more precisely the two-body energy could change it. It is nonetheless natural to assume that the nonlinear radius of the full two-body problem is the one associated to the total mass $M$, i.e.,
\begin{equation}
r_* = \left( \frac{27}{4} \right)^{1/4} \sqrt{M} \; ,
\end{equation}
knowing that the real two-body Vainshtein radius (defined as the radius of convergence of the energy expansion for $r \rightarrow \infty$) could differ by numerical factors dependant on $x$.

Alternativaly, we can formulate the equivalent one-body problem in a different way which will prove useful when investigating the behavior inside the nonlinear radius. Using the energy of a test-mass $\mu$ in an external field generated by $M=m_1+m_2$,
\begin{align} \label{eq:pp_energy}
\begin{split}
\frac{E_\mathrm{tm}}{4 \pi} &= \mu \phi_M(r) \\
 &= \mu \sum_{N \geq 0} \alpha_N \frac{M^{2N+1}}{r^{4N+1}} \; ,
\end{split}
\end{align}
we can build an \textit{energy map} between the real energy $E$ and the effective test-mass energy $E_\mathrm{tm}$ as follows
\begin{align} \label{eq:energy_map_outside}
\begin{split}
\frac{E}{E_\mathrm{tm}} &= f\left(\frac{E_\mathrm{tm}}{E_N}-1\right) \\
&= a_0 + a_1 \left(\frac{E_\mathrm{tm}}{E_N}-1\right) + a_2 \left(\frac{E_\mathrm{tm}}{E_N}-1\right)^2 + \dots
\end{split}
\end{align}

Here $E_N = - \mu M/r$ is the Newtonian reference energy, and the function $f$ has been Taylor expanded for small values of the dimensionless ratio $E_\mathrm{tm}/E_N-1$. Indeed, from eq. \eqref{eq:spherical_symmetry} this ratio can be expanded outside the nonlinear radius as
\begin{equation}
\frac{E_\mathrm{tm}}{E_N}-1 = - \frac{M^2}{5r^4} + \dots
\end{equation}

To construct such an energy map, one can choose each value $a_N$ such that each coefficient in front of $(M^2/r^4)^N$ of eq. \eqref{eq:energy_map_outside} matches. Since the small ratio $E_\mathrm{tm}/E_N-1$ is chosen such that $a_N \left(E_\mathrm{tm}/E_N-1 \right)^N$ contributes only at order $(M^2/r^4)^N$ or higher, this procedure yields an unambiguous value for $a_N$ for all $N$.

This energy map proves very useful because it allows to resum the nonlinear behavior into the small parameter $E_\mathrm{tm}/E_N-1$. We refer the reader to Ref. \cite{buonanno_effective_1999} for its derivation in the context of the Post-Newtonian formalism.

\subsection{Energy map inside}

Having understood the behavior of the energy outside the nonlinear radius, we would like now to generalize to astrophysical situations of interest where the two bodies lie deep within their nonlinear radius. The above formula \eqref{eq:energy_outside_exact} for the two-body energy, even if it resums part of the nonlinear corrections, has no chance to be valid inside $r_*$ because it contains the Newtonian reference energy \eqref{eq:Newtonian_energy} valid only at large radius. However, one can still obtain the coefficients of an energy map from numerical simulations. In App. \ref{sec:matching} we adopt an analytical approach that attempts to relate the two energy maps by a matching condition at the nonlinear radius. While we will argue that this part should yield a qualitative result concerning the two-body energy, we will see when comparing to the numerical simulation that even this qualitative result does not compare well to the real two-body energy. Further improvement is needed in order to obtain a sensible analytical result.

In the following, we will \textit{assume} that the two-body energy for $r<r_*$ can be related, in a spirit similar to the one for $r>r_*$, to the energy of a test-mass $\mu$ in an external field generated by the total mass $M$ through an energy map. We have no possibility to calculate directly the modified coefficients as we did in the last Section, but we can obtain them numerically as we will do in Sec. \ref{sec:numerical}.

Inside the nonlinear radius, the energy map should take the form
\begin{align} \label{eq:energy_map_inside}
\begin{split}
\frac{E'}{E'_\mathrm{tm}} &= g\left(\frac{E'_\mathrm{tm}}{E'_\mathrm{ref}}-1\right) \\
&= b_0 + b_1 \left(\frac{E'_\mathrm{tm}}{E'_\mathrm{ref}}-1\right) + b_2 \left(\frac{E'_\mathrm{tm}}{E'_\mathrm{ref}}-1\right)^2 + \dots
\end{split}
\end{align}

A few comments are required here. First, in order to avoid the appearance of an unphysical (mass-dependant) constant in the energy (which we calculated before by assuming its vanishing at infinity, so it cannot also vanish in zero), we chose to write the energy map using the $r$-derivative of the energy $E'$. Second, we use a reference energy level equal to the small-$r$ value of the energy since the Newtonian energy is irrelevant inside the Vainshtein radius,
\begin{equation}
\frac{E'_\mathrm{ref}}{4\pi} = \mu \left( \frac{M}{r^2} \right)^{1/3} \; .
\end{equation}

Finally, note that the small parameter $E'_\mathrm{tm}/E'_\mathrm{ref} - 1$ is this time expanded as
\begin{align}
\begin{split}
\frac{E'_\mathrm{tm}}{E'_\mathrm{ref}}-1 &= \sum_{N \geq 1} \gamma_N \left( \frac{r^2}{M} \right)^{2N/3} \\
&= - \frac{1}{3} \left( \frac{r^2}{M} \right)^{2/3} + \dots \; ,
\end{split}
\end{align}
where the coefficients $\gamma_N$ can be found by solving exactly the nonlinear equation \eqref{eq:cubic_eq} (we will not need their exact expression here).

In the test-mass limit, the real two-body energy should be approximated by the point-particle energy, and consequently $b_0 = 1$ and $b_1$, $\dots$ $b_N=0$. Now, away from the test-mass limit, we are only interested by the behavior of $b_0$ as a function of the mass ratio $x$, since we recall that the next order is further Vainshtein suppressed and thus irrelevant for astrophysical systems. We will now turn to a numerical implementation that will allow us to obtain the coefficient $b_0$.


\section{Numerical simulation} \label{sec:numerical}

In this Section, we will directly solve the nonlinear PDE in the two-body case using a finite element solver and obtain the two-body energy in order to compare it to the analytical result.

\subsection{Setup} \label{sec:setup}

Using the action \eqref{eq:k_mouflage_rescaled}, one has the following equation of motion for the scalar field
\begin{equation}
\nabla \cdot \left( \nabla \phi + (\nabla \phi)^2 \nabla \phi \right) = -T \; ,
\end{equation}
where we recall that
\begin{equation}
T = -4\pi m_1 \delta^3(\mathbf{x} - \mathbf{x}_1) - 4 \pi m_2 \delta^3(\mathbf{x} - \mathbf{x}_2) \; .
\end{equation}


To numerically solve this equation, a finite element (FEM) solver \footnote{We use the Python FEM solver fenics \cite{ans20553}} is well adapted to the problem, since the PDE can easily be put into a weak form : for any test function $v$ that vanish on the boundary of the integration domain,
\begin{equation} \label{eq:weak_equation}
\int d^3x \left( (1+ (\nabla \phi)^2) \nabla \phi \cdot \nabla v - Tv \right) = 0 \; .
\end{equation}

FEM solvers solve the weak form equation by decomposing the unknown $\phi$ on some basis functions $\psi_j$, here chosen to be the continuous Lagrange polynomials of second order on the grid chosen to discretize the problem. The solver then finds the coefficients $c_j$ of this decomposition $\phi = \sum_j c_j \psi_j$ by evaluating the weak form equation \eqref{eq:weak_equation} with the test function $v$ being one of the basis functions $\psi_i$. This produces a matrix equation for the unknown vector of coefficients $(c_j)_{j \geq 0}$ that is solved by an efficient sparse LU decomposition. The non-linear term is dealt with Newton iterations, i.e. by setting $\phi = \phi_0 + \phi_1$ with $\phi_0$ a function that is close to the solution sought after, linearizing over $\phi_1$ then solving for it, and finally iterate the procedure until a desired convergence threshold has been reached.

In cylindrical coordinates (so that the problem becomes effectively two-dimensional), we choose the two bodies to lie along the $z$ axis at positions $+a/2$ and $-a/2$. We regularize the delta-functions by replacing them with Gaussians,
\begin{align}
\begin{split}
\delta^3(\mathbf{x}) &= \frac{\delta(r) \delta(z)}{2\pi r} \\
&= \frac{1}{2\pi^2 \sigma^2 r} e^{-\frac{r^2+z^2}{2\sigma^2}} \; ,
\end{split}
\end{align}
where we have taken care of the fact that the $r$ variable goes from $0$ to $\infty$ while $z$ goes from $- \infty$ to $\infty$.

There are two scales involved in this problem, the separation between the two bodies that we denote by $a$ and the nonlinear scale $r_* = \sqrt{M}$. We choose for the domain of integration the half-disk defined by $r>0$, $R \leq R_\mathrm{max}$ where $R = \sqrt{r^2+z^2}$. The domain is automatically discretized by the FEM solver with a resolution of approximately $64 \times 64$ points, with a manual refinement of the grid near the two bodies. The boundary conditions are chosen such that $\partial_r \phi (0,z) = 0$ (as required by the symmetry of the problem) and $\phi(r,z) = - \frac{m_1+m_2}{R}$ for $R=R_\mathrm{max}$. The second boundary condition corresponds to recovering both spherical symmetry and the Newtonian-like behavior of the field far from the two bodies, where we know the exact solution which is given by eq. \eqref{eq:spherical_symmetry}. For it to be consistent, we must also ensure $r_* \ll R_\mathrm{max}$.

Once we get the field solution, the energy can be computed as
\begin{align}
\begin{split}
E &= \int d^3x \left( \frac{(\nabla \phi)^2}{2} +  \frac{(\nabla \phi)^4}{4} \right) \\
&+ 4\pi m_1 \phi(\mathbf{x}_1) + 4\pi m_2 \phi(\mathbf{x}_2) \; .
\label{eq:numerical_energy}
\end{split}
\end{align}

A remarkable fact to be noted is that the self-energy contribution $4\pi m_1 \phi(\mathbf{x}_1) + 4\pi m_2 \phi(\mathbf{x}_2)$ is not divergent, contrary to the Newtonian case. This is due to the fact that the field goes to a constant as $\phi \sim |\mathbf{r} - \mathbf{x}_\alpha|^{1/3}$ close to the source $\alpha$, instead of diverging as $\frac{1}{|\mathbf{r} - \mathbf{x}_\alpha|}$. The same goes for the integral over all space. We consequently do not need to renormalize the energy.

Finally, after having compared the behavior of the energy for $r \gtrsim r_*$ to the theoretical predictions, we will need the energy in the realistic $r \ll r_*$ case. In this setup, we can ignore the quadratic term in the equation of motion \eqref{eq:weak_equation}. This corresponds to take an infinite Vainshtein radius. Correspondingly, we have to change the exterior boundary condition to $\phi(r,z) = 3((m_1+m_2)R)^{1/3}$ for $R=R_\mathrm{max}$ because the field is screened over all space. The energy is computed ignoring also the quadratic term, but there is a subtelty in its definition because the integral is formally divergent as $R_\mathrm{max}^{1/3}$ (a consequence of the good UV and bad IR behavior of the field). This divergence is the same as the one associated to a single particle of mass $M=m_1+m_2$ sitting at the origin, and so to renormalize the theory we susbtract to the energy the contribution of such a field which is
\begin{equation}
3\pi ((m_1+m_2)R_\mathrm{max})^{1/3} \; .
\end{equation}



\subsection{Numerical tests} \label{sec:num_tests}

In order to assess the validity of our numerical scheme, we have performed two numerical tests. Before presenting them, let us first discuss the choice of the numerical parameters $R_\mathrm{max}$ and $\sigma$. A finite choice of $R_\mathrm{max}$ brings a correction of the order $a/R_\mathrm{max}$ to the field (where we recall that $a$ is the separation between the two bodies), and so we choose $R_\mathrm{max}=50a$  in our simulation.

Concerning the effect of $\sigma$, let us consider a single point-particle of mass $m_2$ in the $r \ll r_*$ case with boundary condition $\phi(r,z) = 3(m_2 R)^{1/3}$ for $R=R_\mathrm{max}$. In this case, if the particle was really pointlike, the field would vanish at the origin. This is no longer the case if the particle has a finite extent $\sigma$. Rather, the central value of the field is
\begin{equation} \label{eq:field_sigma}
\phi(0,0) \sim (m_2 \sigma)^{1/3} \; ,
\end{equation}
simply by dimensional analysis. But this could potentially be problematic in the calculation of the two-body energy which requires the evaluation of the field at the particle location, see eq. \eqref{eq:numerical_energy}. The contribution of this term in the total two-body energy is
\begin{align}
\begin{split}\label{eq:field_center}
m_2 \phi(\mathbf{x}_2) &\sim E_\mathrm{tm} \\ 
&\sim \frac{m_1m_2}{m_1+m_2} ((m_1+m_2)a)^{1/3} \; ,
\end{split}
\end{align}
this equality being true up to a numerical factor.

In order to avoid an unphysical dependance on $\sigma$ in the energy, we have to tune $\sigma$ in order that the term in eq. \eqref{eq:field_sigma} is much smaller than the term in eq. \eqref{eq:field_center}. This gives the following bound:
\begin{equation} \label{eq:condition_sigma}
\sigma \ll \frac{x^3}{1-x} a \; ,
\end{equation}
where we recall that $x = m_1/(m_1+m_2)$ is the mass ratio. We choose $\sigma = 10^{-6} a$, which is consistent with the minimal value of $x$, $x_\mathrm{min}=0.03$ for $a=1$, that we will use.

\begin{figure}
\includegraphics[width=\columnwidth]{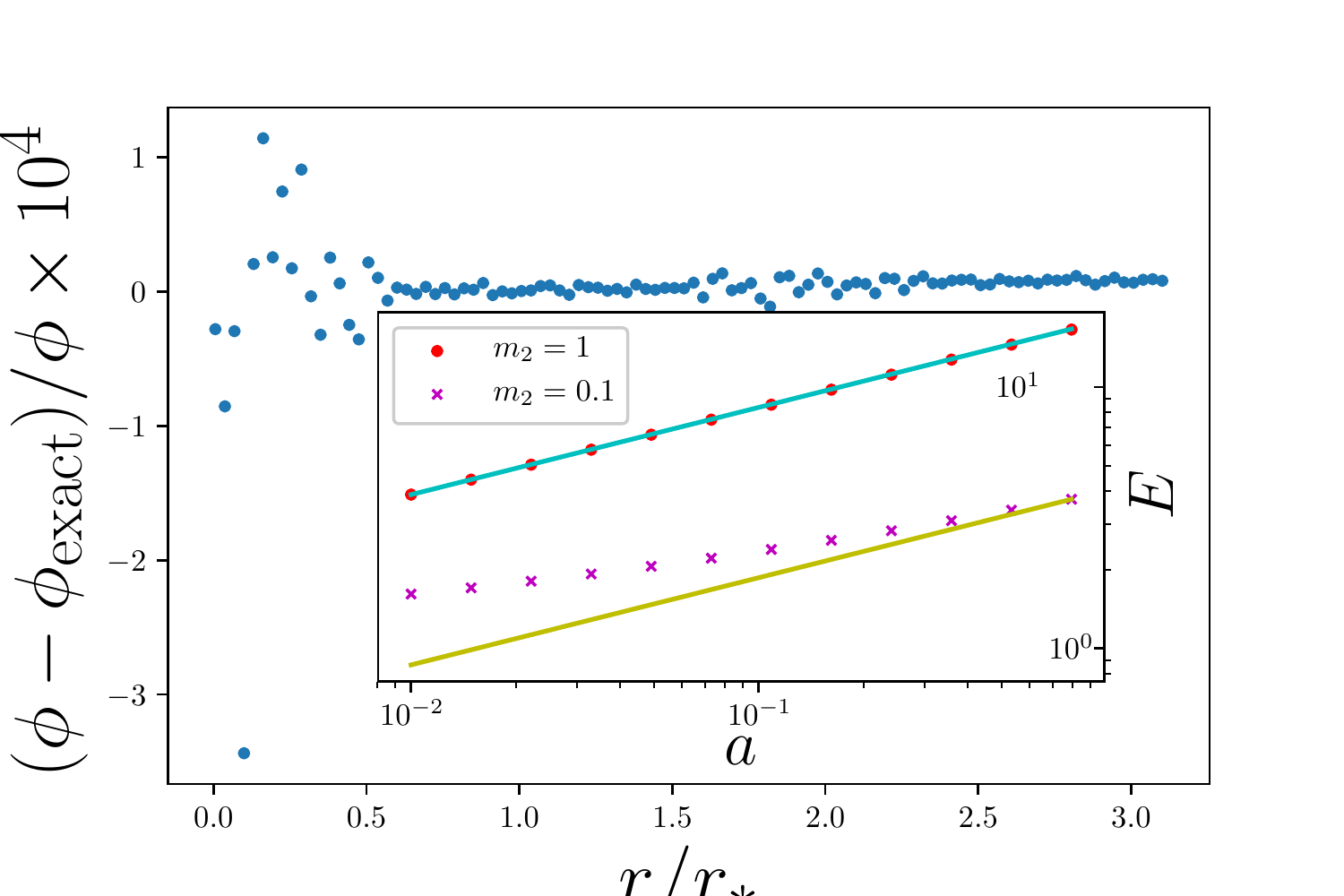}
\caption{Checks of the numerical code. The main plot is the fractional difference of the numerical field solution in the spherically symmetric case of a single particle of mass $M = 1$, compared to the exact solution of eq. \eqref{eq:hypergeom}.
The subplot is the two-body energy defined by eq. \eqref{eq:numerical_energy} in the $r \ll r_*$ case, keeping $m_1=1$ and for two values of $m_2$. The parameter $\sigma$ is taken to be $\sigma = 10^{-4}$. The upper points, with parameter $m_2 = 1$, show good agremment between the expected power-law behavior $E \propto a^{1/3}$ (continuous cyan curve) and the numerical result. The agreement between the two is a good check of the validity of our code. The bottom points, with $m_2 = 0.1$, show a deviation from the simple power-law behavior (continuous yellow curve) due to the fact that condition \eqref{eq:condition_sigma} is not satisfied any more.}
\label{fig:spher_sym}
\end{figure}

Having chosen a value for the numerical parameters, we have first checked that, for a single particle at the origin of the coordinates, the field solution corresponds to the exact solution given in eq. \eqref{eq:hypergeom}. The result is plotted in Figure \ref{fig:spher_sym}, where we see that there is perfect agreement between the theoretical and numerical values.

The second nontrivial check that we have performed is to verify that the two-body energy in the fully screened regime (i.e, neglecting the quadratic operator as explained in Sec. \ref{sec:setup}) indeed varies as $E \propto a^{1/3}$. In Figure \ref{fig:spher_sym} we have plotted the numerical value of the two-body energy in the $r \ll r_*$ case in logarithmic plot, from which we immediately confirm the expected behavior.


\subsection{Results} \label{sec:results}

\begin{figure}
\includegraphics[width=\columnwidth]{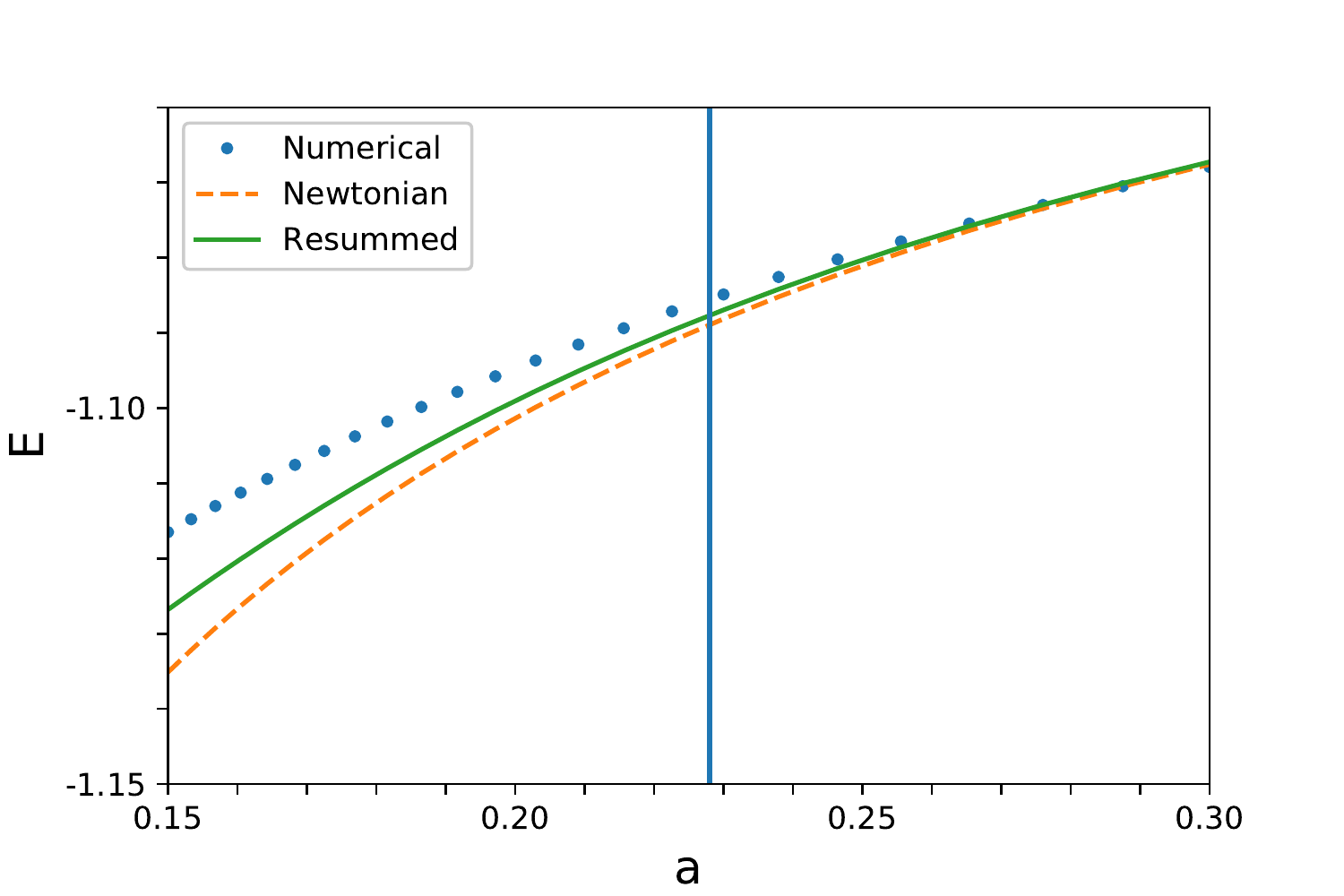}
\caption{Plot of the numerical solution to the energy close to the Vainshtein radius as a function of the point-masses spacing $a$, for parameters $m_1=m_2=10^{-2}$. The energy is normalized to its absolute value at infinity \eqref{eq:E_infinity}. The numerical solution corresponds to the blue filled circles, the Newtonian potential $-m_1m_2/r+E_\infty$ to the dashed line and the resummed solution \eqref{eq:energy_outside_exact} (shifted  with respect to $E_\infty$) to the solid curve. The vertical bar is the location of the Vainshtein radius associated to the total mass $M=m_1+m_2$.}
\label{fig:energy_outside}
\end{figure}

Figure \ref{fig:energy_outside} presents the numerical two-body in the $r \gtrsim r_*$ regime against different theoretical predictions for equal masses. The value of the energy when the charges are taken infinitely far apart is not zero: it can be easily calculated as two times the value of the energy when we plug the exact one-body field solution \eqref{eq:hypergeom} into the action. This yields
\begin{equation}\label{eq:E_infinity}
E(\infty) \simeq -31 (m_1^{3/2}+m_2^{3/2}) \; .
\end{equation}

We can see that the numerical value of the energy for large separation matches very well this result, thus providing again a strong check of the validity of our code. The resummed energy provides an improvement over the simple Newtonian potential, although the gain is $\sim 50$ \% at the nonlinear radius.

\begin{figure}
\includegraphics[width=\columnwidth]{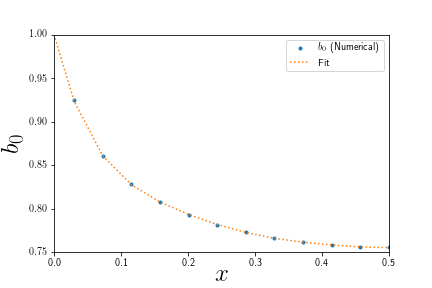}
\caption{Plot of the numerical solution for the coefficient $b_0$ defined in eq. \eqref{eq:b0_num}, for a spacing between point-particles of $a=1$.}
\label{fig:b0_num}
\end{figure}

Concerning the fully screened situation $r \ll r_*$, Figure \ref{fig:b0_num} presents the effective parameter $b_0$ as a function of the mass ratio $x$, where $b_0$ is defined as
\begin{align} \label{eq:b0_num}
\begin{split}
b_0 &= \frac{E}{E_\mathrm{tm}} \; , \\
\frac{E_\mathrm{tm}}{4\pi} &= 3 \mu (Ma)^{1/3} \; ,
\end{split}
\end{align}
where we chose $a=1$ to obtain our results ($b_0$ does not depend on $a$, as emphasised in Sec. \ref{sec:num_tests}) and the numerical energy $E$ is computed by keeping only the nonlinear term in the action, as explained in Sec. \ref{sec:setup}. This is the most important result of this work, since it presents the two-body energy in astrophysically relevant situations and for arbitrary mass ratios. We can see that $b_0 \simeq 0.75$ in the equal-mass case, which means that screening is a bit more efficient than when there is a large mass hierarchy (for which $b_0 = 1$). For convenience, a fit to $b_0$ with a sixth-order polynomial gives

\begin{align}
\begin{split}
b_0(x) &= 1 -3.17 x + 23.7 x^2 - 105.89 x^3 \\
 &+ 266.48 x^4 - 347.46 x^5 + 182.64 x^6
\end{split}
\end{align}

The dependance of the two-body energy on the mass ratio implies a direct violation of the Weak Equivalence Principle (WEP). In the next Section we will explore the consequences of this result on the orbit of the Moon, which is measured with a great accuracy by Lunar Laser Ranging.

\section{Violation of the Weak Equivalence Principle} \label{sec:WEP}

In this Section we will show how our results imply an Equivalence Principle violation that would be visible on the Moon orbit. We will make heavy use of the following mass ratios
\begin{align}
\begin{split}
x_\mathrm{SE} = \frac{m_\oplus}{m_\odot + m_\oplus} &\simeq 3 \times 10^{-6} \\
x_\mathrm{SM} = \frac{m_M}{m_\odot + m_M} &\simeq 3 \times 10^{-8} \\
x_\mathrm{EM} = \frac{m_M}{m_\oplus + m_M} &\simeq 10^{-2} \; ,
\end{split}
\end{align}
where $m_\odot$ is the Sun mass, $m_\oplus$ is the Earth mass and $m_M$ is the Moon mass. 

\subsection{Three-body system, and finite size corrections} \label{sec:three-body}

In this part we will examine the applicability of our results to a three body system like the one formed by the Sun, the Earth and the Moon (herafter, SEM system). Fifth forces generated by the scalar interaction are expected to induce a supplementary perihelion precession that would be visible on planetary orbits \cite{PhysRevLett.61.1159}. But the lunar perihelion precession cannot be computed by ignoring the scalar field generated by the Sun, and using a perturbative treatment Ref. \cite{andrews_galileon_2013} reached the conclusion that the perturbations blows up once distances hierarchies are taken into account. We will recover their result using a different approach.


To start with, let us reformulate the action of a two-body system in a different way. Labelling the two objects as $1$ and $2$, we split the scalar field according to
\begin{equation}
\phi = \phi_2(r_2) + \psi \; ,
\end{equation}
where $r_2 = \vert \mathbf{x} - \mathbf{x}_2 \vert$ is the distance to the body $2$, and $\phi_2$ is the spherically symmetric field \eqref{eq:spherical_symmetry} generated by the same body. In the fully screened situation that is of interest to us, this fields writes as 
\begin{equation}
\phi_2(r_2) = 3 (m_2 r_2)^{1/3} \; .
\end{equation}

If $m_2$ is much greater than the other mass, we can think of $\psi$ being a fluctuation on top of the dominant field generated by the mass $m_2$, but for now let's keep the discussion general and not assume any mass hierarchy (we still assume $m_1 \leq m_2$ by symmetry). Then by inserting this decomposition in the action \eqref{eq:k_mouflage_rescaled} (where we ignored the quadratic term that is subdominant on small scales), we get the following action
\begin{align}
\begin{split} \label{eq:expanded_action}
S &= S[\phi_2] + \Lambda^4 \int dt d^3x \left[ -\frac{1}{2} (\nabla \phi_2)^2 (\nabla \psi)^2 - (\nabla \phi_2 \cdot \nabla \psi)^2  \right. \\
&- \left. (\nabla \psi)^2 \nabla \phi_2 \cdot \nabla \psi - \frac{1}{4} (\nabla \psi)^4 + \psi T_1 \right] \; ,
\end{split}
\end{align}
where $S[\phi_2]$ is the original action \eqref{eq:k_mouflage_rescaled} applied to $\phi_2$, and $T_1 = -4\pi m_1 \delta^3(\mathbf{x} - \mathbf{x}_1)$ is the source term corresponding to the object $1$. The term linear in $\psi$ vanishes because of the equations of motion for $\phi_2$.

We now ask the question : close to the object $1$, what is the behavior of the fluctuation $\psi$ ? In other words, is there an operator that dominates the action for $\psi$ in \eqref{eq:expanded_action} ? It seems natural to assume that close enough to the source $1$, we recover the behavior $\psi \sim (m_1 r_1)^{1/3}$, which means that the last nonlinear operator in eq. \eqref{eq:expanded_action} dominates. Let us assume this is the case and derive the condition on $r_1$ for this to be true.

If the term $(\nabla \psi)^4$ dominates in the action, then we recover the same spherically symmetric action as for the second object, and we consequently obtain
\begin{equation}
\psi \simeq 3 (m_1 r_1)^{1/3} \; .
\end{equation}

Let us now make the ratio between the operator $(\nabla \psi)^4$ and another one in the action, say the term cubic in $\psi$. Using Cauchy-Schwarz inequality, we get
\begin{equation} \label{eq:PX_condition_background}
\frac{(\nabla \psi)^2 \nabla \phi_2 \cdot \nabla \psi}{(\nabla \psi)^4} \leq \frac{\vert \nabla \phi_2 \vert}{\vert \nabla \psi \vert} \simeq \left( \frac{m_2}{m_1} \left(\frac{r_1}{r_2} \right)^2 \right)^{1/3} \; .
\end{equation}

Using the same scaling, one can show that the term quadratic in $\psi$ is similarly suppressed with respect to the term cubic in $\psi$. This result is quite important. It means that around the first body, the field can be well approximated simply by taking the linear superposition $\phi_2 + \psi$ of two spherically symmetric solutions. This is true up to the maximal distance to the first body
\begin{equation}
r_1^\mathrm{max} = r_2 \sqrt{\frac{m_1}{m_2}} \;,
\end{equation}
which depends on the mass ratio. Since the Vainshtein radius of a massive body is $r_{*, \alpha} = \sqrt{m_\alpha}$, this equation can also be interpreted as
\begin{equation}
\frac{r_1^\mathrm{max}}{r_{*, 1}} = \frac{r_2}{r_{*,2}}
\end{equation}
which simply means that the distance to each body is measured in units of its Vainshtein radius.

This has several consequences. First, it means that we can ignore the finite-size of the bodies and treat them as point-particles as long as their radius is less than $r_1^\mathrm{max}$. In this particular theory, this is true both for the Sun-Earth system and for the Earth-Moon system if we take the numerical values of their respective radius and masses.

Second, it means that we can simply add the fifth forces felt by a satellite that orbits sufficiently close to a planet itslef orbiting around its star. Unfortunately, for the SEM system and the particular screening theory considered here this is not true, as
\begin{equation}
\left( \frac{m_\odot}{m_\oplus} \left(\frac{r_1}{r_2} \right)^2 \right)^{1/3} \simeq 1 \; ,
\end{equation}
where we have taken $r_1$ to be the Earth-Moon distance and $r_2$ to be the Sun-Earth distance. The end result is that we expect the Lunar perihelion precession rate to be corrected by an amount depending on the masses and distances hierarchies.

For another type of Vainshtein screening such as the Galileon-3 \cite{nicolis_galileon_2009}, the same reasoning shows that we can superpose the nonlinear solutions provided
\begin{equation} \label{eq:galileon_condition_background}
\left( \frac{m_2}{m_1} \left(\frac{r_1}{r_2} \right)^3 \right)^{1/2} \ll 1 \; ,
\end{equation}
which is true at the 10\% level for the SEM system. The same conclusion was also reached in \cite{Nicolis:2004qq} with similar arguments. This shows that in the case of a Galileon-3, the Lunar perihelion precession can be calculated by simply ignoring the Sun \footnote{Note that we can also apply this reasoning to the Sun embedded inside our own galaxy. In this case, $m_2/m_1 = m_\mathrm{Gal}/m_\odot \simeq 10^{12}$, $r_1/r_2 = r_{\mathrm{SE}}/r_\mathrm{Gal} \simeq 6 \times 10^{-10}$ where $m_\mathrm{Gal}$ is the galactic mass, $r_\mathrm{Gal}$ is the distance to the Galactic centre, and we have made the simplifying assumption that the mass of the Galaxy is concentrated at its centre. Eqs \ref{eq:PX_condition_background} and \ref{eq:galileon_condition_background} then show that, for both Galileons and $P(X)$, we can indeed neglect the background field generated by the galaxy for the inner planets of the Solar System. I am grateful to Philippe Brax for pointing this out.} Since the Earth-Moon mass ratio is $x \simeq 10^{-2}$, the calculation can be carried out in the test-mass approximation and yields an interesting constraint on the size of a Galileon-3 operator \cite{dvali_accelerated_2003}.


\subsection{Weak Equivalence Principle violation}


The fact that the two-body energy of two massive particles is not the one of a reduced mass $\mu$ in an external field implies a violation of the Weak Equivalence Principle, as we will now show. In this Section we will assume a general Vainshtein screening mechanism that gives rise to a fifth force (not necessarily the specific $P(X)$ theory that we studied in the rest of the article). If we first neglect the moon, the total (gravitational and scalar) interacting Lagrangian of the Earth and the Sun writes as
\begin{equation} \label{eq:total_2_body_lagrangian}
L_{SE} = \frac{G m_\odot m_\oplus}{r_{SE}} \left(1 + \alpha(x_{SE}) \left(\frac{r_{SE}}{r_*} \right)^n \right) \; .
\end{equation}
Here $r_{SE} = \vert \mathbf{y}_S - \mathbf{y}_E \vert$ is the Earth-Sun distance, $x_{SE} = m_\oplus/(m_\odot + m_\oplus)$ is the Earth-Sun mass ratio, $r_*$ is the Vainshtein radius of the Sun, $n$ is an exponent that depends on the type of screening considered, and $\alpha$ is an unknown function of the coefficient $x_{SE}$ that can be found numerically as in Sec. \ref{sec:numerical}. In order to avoid confusion between positions and mass ratios, we denote in this Section the positions by the letter $\mathbf{y}$. This type of Lagrangian is common to all theories endorsed with Vainshtein screening, with the expressions of $n$, $r_*$ and $\alpha$ different among theories. For example, $n = 4/3$ for the $P(X)$ theory considered above in this article, and $n = 3/2$ for a Galileon-3.

We will make the supplementary assumption that we can get the total force felt by the Moon by simply adding the fifth forces of the Earth and the Sun. As we showed in Sec. \ref{sec:three-body}, this is the case for a Galileon-3 but not for the particular $P(X)$ example examined in the rest of the article. A complete treatment of this case would necessitate further work. With this assumption we can write the total (non-relativistic) interaction Lagrangian of this three-body system (to get the total Lagrangian from this, one should also add the kinetic energies from the three bodies)
\begin{align} \label{eq:total_3_body_lagrangian}
\begin{split}
L_\mathrm{int} &= \frac{G m_\odot m_\oplus}{r_{SE}} \left(1 + \alpha(x_{SE}) \left(\frac{r_{SE}}{r_*} \right)^n \right) \\
&+ \frac{G m_\odot m_M}{r_{SM}} \left(1 + \alpha(x_{SM}) \left(\frac{r_{SM}}{r_*} \right)^n \right) \\
&+ \frac{G m_\oplus m_M}{r_{EM}} \left(1 + \alpha(x_{EM}) \left(\frac{r_{EM}}{r_*} \right)^n \right) \; ,
\end{split}
\end{align}
where $S$ designates the Sun, $E$ the Earth, $M$ the Moon, and each line of this equation is the two-body Lagrangian of eq. \eqref{eq:total_2_body_lagrangian} adapted to each pair of bodies. We will now derive the fifth force incidence on the Lunar motion along the lines of Ref. \cite{Damour:1995gi}.

The first line of eq. \eqref{eq:total_3_body_lagrangian} gives rise to the Earth anomalous perihelion precession, and the third line to the Lunar anomalous perihelion precession. These effects are already discussed in other References \cite{dvali_accelerated_2003, Iorio:2012pv, andrews_galileon_2013} and we will not comment on them. From now on, we will ignore the last line of eq. \eqref{eq:total_3_body_lagrangian} which does not give rise to the leading order equivalence principle violation that we are going to derive. Let us expand the distances to the Sun around the Earth-Moon center-of-mass, which is defined with the usual expression
\begin{equation}
(m_\oplus + m_M) \mathbf{Y} = m_\oplus \mathbf{y}_E + m_M \mathbf{y}_M \; ,
\end{equation}
where $\mathbf{y}_E$ and $\mathbf{y}_M$ are the positions of the Earth and the Moon respectively. Then the distances to the Sun can be expressed by
\begin{align}
\begin{split}
r_{SE} &= \vert \mathbf{y}_S - \mathbf{Y} - x_{SM} \mathbf{y}_{EM}  \vert \\
r_{SM} &= \vert \mathbf{y}_S - \mathbf{Y} + x_{SE} \mathbf{y}_{EM}  \vert \; ,
\end{split}
\end{align}
where $\mathbf{y}_S$ is the Sun position. Expanding the total Lagrangian to first order in $r_{EM}$, one finds
\begin{widetext}
\begin{align}
\begin{split}
L_\mathrm{int} &= \frac{G m_\odot (m_\oplus + m_M)}{r} \left(1 + \left[ (1-x_{EM}) \alpha(x_{SE}) + x_{EM} \alpha(x_{SM}) \right] \left(\frac{r}{r_*} \right)^n \right) \\
&- G m_\odot m_\oplus x_{EM} r_{EM}^i \frac{\partial}{\partial r^i} \left[ \frac{1}{r} \left(1+\alpha(x_{SE}) \left(\frac{r}{r_*} \right)^n  \right) \right]
+ G m_\odot m_M (1-x_{EM}) r_{EM}^i \frac{\partial}{\partial r^i} \left[ \frac{1}{r} \left(1+\alpha(x_{SM}) \left(\frac{r}{r_*} \right)^n  \right) \right] \;,
\end{split}
\end{align}
where $r = \vert \mathbf{y}_S - \mathbf{Y} \vert$ is the center-of-mass distance to the Sun, and as discussed above we have dropped the third line of eq. \eqref{eq:total_3_body_lagrangian}. Upon introducing the reduced mass $\mu_{EM} = m_\oplus m_M / (m_\oplus + m_M)$, one finds the following expression
\begin{align} \label{eq:reduced_3_body_lagrangian}
\begin{split}
L_\mathrm{int} &= \frac{G m_\odot (m_\oplus + m_M)}{r} \left(1 + \left[ (1-x_{EM}) \alpha(x_{SE}) + x_{EM} \alpha(x_{SM}) \right] \left(\frac{r}{r_*} \right)^n \right) \\
&+ G \mu_{EM} m_\odot \frac{r_{EM}^i r_i}{r^3} \left( (1-n) (\alpha(x_{SE})-\alpha(x_{SM})) \left(\frac{r}{r_*} \right)^n \right) \; .
\end{split}
\end{align}
\end{widetext}

As discussed in Ref. \cite{Damour:2012mv}, there are two physical effects that stem from this Lagrangian \footnote{In fact Ref. \cite{Damour:2012mv} discusses three physical consequences, but the third one is not apparent at the expansion order we work with, and is expected not to be measureable.}. The first line of eq. \eqref{eq:reduced_3_body_lagrangian} implies that the gravitational constant involved in the motion of the Earth-Moon system around the Sun is not the same than the one involved in the motion of this system around its barycenter (third line of eq. \eqref{eq:total_3_body_lagrangian}). However, this effect has practically no observable consequences.

The second physical effect, on which we will concentrate, is the perturbation on the lunar orbit implied by the second line of eq. \eqref{eq:reduced_3_body_lagrangian}. Nordtvedt \cite{PhysRev.170.1186} showed in 1968, in the context of Scalar-Tensor theories, that this term implies a modulation of the Lunar orbit with amplitude
\begin{equation}
\delta r_{EM} \simeq 3 \times 10^{12} \vert \delta_\oplus - \delta_M \vert \; \mathrm{cm} \; .
\end{equation}
In scalar-tensor theories, $\delta_{AB}$ is the fractional variation of Newton's constant due to the gravitational self-energy of each body ; said equivalently, Newton's constant appearing in the gravitational attraction between two bodies is
\begin{equation}
G_{AB} = G(1+\delta_A + \delta_B) \; .
\end{equation}

Although the physical origin is quite different, the equivalence principle violation considered here gives rise to the same term in the second line of the Lagrangian \eqref{eq:reduced_3_body_lagrangian} than the equivalence principle violation of Scalar-Tensor theories. The parameter $\delta_\oplus - \delta_M$ is replaced by the following quantity
\begin{align}
\begin{split}
\delta_\oplus - \delta_M &\rightarrow (1-n) (\alpha(x_{SE})-\alpha(x_{SM})) \left(\frac{r}{r_*} \right)^n \\
& \simeq (1-n) \alpha_1 x_{SE} \left(\frac{r}{r_*} \right)^n \; .
\end{split}
\end{align}
In the second line we have expanded $\alpha$ to first order in the mass ratios, $\alpha(x) \simeq \alpha_0 + \alpha_1 x$ and neglected the Moon mass ratio compared to the Earth mass ratio, $x_{SM} \ll x_{SE}$.

Current LLR data give the constraint $\vert \delta_\oplus - \delta_M \vert \lesssim 10^{-13}$ \cite{Williams:2012nc}, which by ignoring the $\mathcal{O}(1)$ factor of $(1-n)$ yield the constraint
\begin{equation}
\alpha_1 x_{SE} \left(\frac{r}{r_*} \right)^n \lesssim 10^{-13}
\end{equation}

Let us contrast this with the anomalous perihelion constraint. For the earth, the precession is measured with an accuracy of $10^{-11}$ (for a precise constraint on a Galileon-3, see Ref. \cite{Iorio:2012pv}), and the anomalous perihelion precession in typical Vainshtein screened theories is proportional to the ratio of scalar and gravitational potentials \cite{dvali_accelerated_2003}, which gives the constraint
\begin{equation}
\alpha_0 \left(\frac{r}{r_*} \right)^n \lesssim 10^{-11}
\end{equation}

On the one hand, we gain two orders of magnitude by using Lunar Laser Ranging data, but on the other hand the equivalence principle violation is suppressed by the mass ratio $x_{SE} \simeq 10^{-6}$ with respect to the perihelion bound, resulting in a looser constraint if we assume that $\alpha_0 \sim \alpha_1 \sim \mathcal{O}(1)$.

\section{Conclusions} \label{sec:conclusion}

In this paper we have analyzed for the first time the two-body potential energy of pointlike objects in Vainshtein screened theories for arbitrary mass ratios. One the one hand, from outside the nonlinear radius, the problem is amenable to a perturbative treatment which we use to resum a class of Feynman graphs. We derive an Effective One-Body energy map which relates the two-body energy to the one of a test particle in an external field. On the other hand, we conjecture the existence of such an energy map inside the nonlinear radius where the nonlinear screening term dominates the action. We have tried to get the analytical behavior of this expansion by a matching procedure at the nonlinear scale. Improving the accuracy of this matching procedure would necessitate to know the exterior potential energy with a higher accuracy. This could be done by calculating Feynman diagrams with \textit{two} insertions of the nonlinear operator, instead of only one as we did in Sec. \ref{sec:two_body}. This would necessitate more involved computations that we leave for future work.

At the same time, we have performed a numerical simulation in order to get the most relevant effective coefficient $b_0$ corresponding to the ratio between the real two-body energy and the energy of a test-mass in an external field. In the test-mass limit, $b_0 \simeq 1$ while in the equal-mass limit, we find $b_0 \simeq 0.75$ in the particular $P(X)$ theory that we considered. This means that Vainshtein screening is active even in the fully nonlinear situation where we do not assume one mass to be smaller than the other, with departure from simple order-of-magnitude estimates encoded into a simple coefficient $b_0$ which can be found with a numerical simulation. While we have focused on a particular model exhibiting nonlinearities, we expect such a feature to be valid in any model endorsed with Vainshtein screening such as Galileons.

As we showed in Sec. \ref{sec:WEP}, the fact that the two-body energy differs from the test-mass one implies a violation of the weak equivalence principle, as the Earth and the Moon would not fall the same way towards the Sun. We used this fact to bound the size of the coupling parameters of any theory relying on Vainshtein screening to hide the effects of a fifth force in the Solar system. Although the final constraint is looser than the one obtained from perihelion precession for a Galileon-3, the methodology employed is quite general and we intend to use it to investigate on the case of two neutron stars or black holes in their inspiral phase.


\begin{acknowledgments}
I would like to thank Federico Piazza and Julien Bel for insightful comments on the draft.
\end{acknowledgments}

\begin{widetext}
\appendix
\section{Useful integrals}
\label{sec:useful_integrals}

\begin{equation}
\int \frac{d^dk}{(2 \pi)^d} \frac{1}{(\mathbf{k}^2)^a ((\mathbf{k} + \mathbf{K})^2)^b} = \frac{1}{(4\pi)^{d/2}} \frac{\Gamma(a+b-d/2) \Gamma(d/2-a) \Gamma(d/2-b)}{\Gamma(a) \Gamma(b) \Gamma(d-a-b)} (\mathbf{K}^2)^{d/2-a-b}
\end{equation}

\begin{equation}
\int \frac{d^dk}{(2 \pi)^d} \frac{k^i}{(\mathbf{k}^2)^a ((\mathbf{k} + \mathbf{K})^2)^b} = \frac{1}{(4\pi)^{d/2}} \frac{\Gamma(a+b-d/2) \Gamma(d/2-a+1) \Gamma(d/2-b)}{\Gamma(a) \Gamma(b) \Gamma(d-a-b+1)} (\mathbf{K}^2)^{d/2-a-b} K^i
\end{equation}

\begin{align}
\begin{split}
\int \frac{d^dk}{(2 \pi)^d} \frac{k^i k^j}{(\mathbf{k}^2)^a ((\mathbf{k} + \mathbf{K})^2)^b} &= \frac{1}{(4\pi)^{d/2}} \frac{\Gamma(a+b-d/2-1) \Gamma(d/2-a+1) \Gamma(d/2-b)}{\Gamma(a) \Gamma(b) \Gamma(d-a-b+2)} (\mathbf{K}^2)^{d/2-a-b} \\
& \times \left( \frac{d/2-b}{2} \mathbf{K}^2 \delta^{ij} + (a+b-d/2-1)(d/2-a+1)K^i K^j \right)
\end{split}
\end{align}

\begin{equation}
\int \frac{d^dk}{(2 \pi)^d} (\mathbf{k}^2)^\alpha e^{i \mathbf{k} \mathbf{r}} = \frac{2^{2\alpha-1}}{\pi^{(d-1)/2}} \frac{\Gamma(\alpha + d/2)}{\Gamma(d/2) \Gamma((3-d)/2-\alpha)} (\mathbf{r}^2)^{-\alpha -d/2}
\label{eq:intK}
\end{equation}

\end{widetext}

\section{Matching interior and exterior solutions}
\label{sec:matching}

In this Appendix, we will attempt to obtain a qualitative analytical behavior of the energy as a function of the mass ratio inside the nonlinear radius. We can use the formula for the energy outside the screening radius that we found in Sec. \ref{sec:resum_TM} as a limiting boundary condition for the energy map inside the screening radius. Since the energy map partly resums the nonlinear behavior, it should allow to get a qualitative description of the coefficient $b_0$.

Using the energy derived in eq. \eqref{eq:energy_outside_exact}, one finds that the left-hand side of the energy map \eqref{eq:energy_map_inside} writes as
\begin{align}
\begin{split}
\frac{E'}{E'_\mathrm{tm}} &= \frac{1}{\phi'_M} \left[ -\frac{M}{r^2} + \frac{\phi'_{M(1-x)}}{1-x} + \frac{\phi'_{Mx}}{x} \right] \\
&\equiv F(r,x) \; ,
\end{split}
\end{align}
for $r$ close to $r_*$. We recall that $x$ is the mass ratio $x = m_1/M \leq 1/2$ and that $\phi_m$ is the spherically symmetric field  \eqref{eq:spherical_symmetry} generated by a mass $m$. Note that, as $\frac{\phi'_{Mx}}{x} \rightarrow \frac{M}{r^2}$ as $x \rightarrow 0$, this expression has the good test-mass limit, and in this limit we recover that $b_0 = 1$ and $b_1, \; \dots \; b_N = 0$.

If the energy map really resums a part of the nonlinear dynamics, then it should give an asymptotic expansion near to the Vainshtein radius, as suggested by the relative smallness of the expansion parameter at $r_*$,
\begin{equation}
\left.\frac{E'_\mathrm{tm}}{E'_\mathrm{ref}}-1 \right|_{r_*} \simeq - 0.53 \; ,
\end{equation}
where we recall that $r_* = \left( 27/4 \right)^{1/4} \sqrt{M}$.

Consequently, we can try to find a first approximation for the value of $b_0$ by neglecting higher-order contributions in the right-hand side of eq. \eqref{eq:energy_map_inside}. This gives for $b_0$
\begin{equation}
b_0^{(0)}(x) \simeq  F(r_*, x) \; .
\end{equation}

\begin{figure}
\includegraphics[width=\columnwidth]{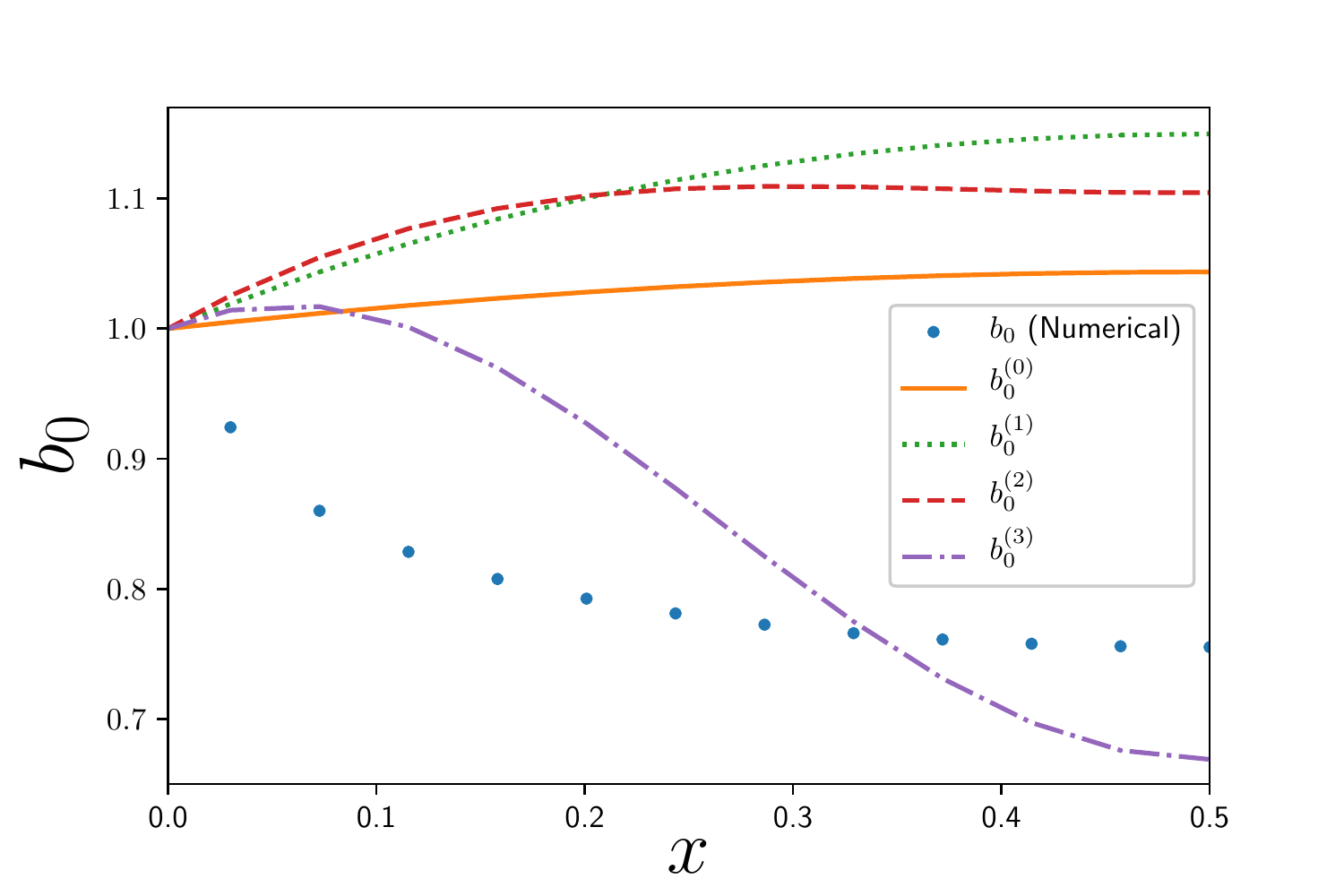}
\caption{Plot of the first four predictions for the coefficient $b_0$ defined in eq. \eqref{eq:energy_map_inside} using the procedure defined in the main text. For comparison, the numerical coefficient $b_0$ is also plotted.}
\label{fig:b0}
\end{figure}

This first prediction for $b_0$ is plotted in Figure \ref{fig:b0}.

One can try to improve the result by matching not only the numerical value of the energy map \eqref{eq:energy_map_inside} at $r=r_*$, but also its derivatives with respect to $r$. This will allow us to extract $b_1$, $b_2$... while at the same time refining our prediction on $b_0$. More precisely, matching up to the $k^\mathrm{th}$ derivative will allow us to determine $k+1$ coefficients in the energy map, say $b_0^{(k)} , \; b_1^{(k)}, \; \dots \; b_k^{(k)}$.

When should we stop this procedure ? Obviously it cannot be pushed to arbitrary order because the boundary condition on the energy is not sufficiently precise. A sensible criterion can be found by thinking in terms of asymptotic series. Let us write the $k^\mathrm{th}$ prediction for $b_0$ as
\begin{equation} \label{eq:asymptotic_series}
b_0^{(k)} = \sum_{l=0}^{k} b_0^{(l)} - b_0^{(l-1)} \; ,
\end{equation}
where we have defined $b_0^{(-1)} = 0$.

\begin{figure}
\includegraphics[width=\columnwidth]{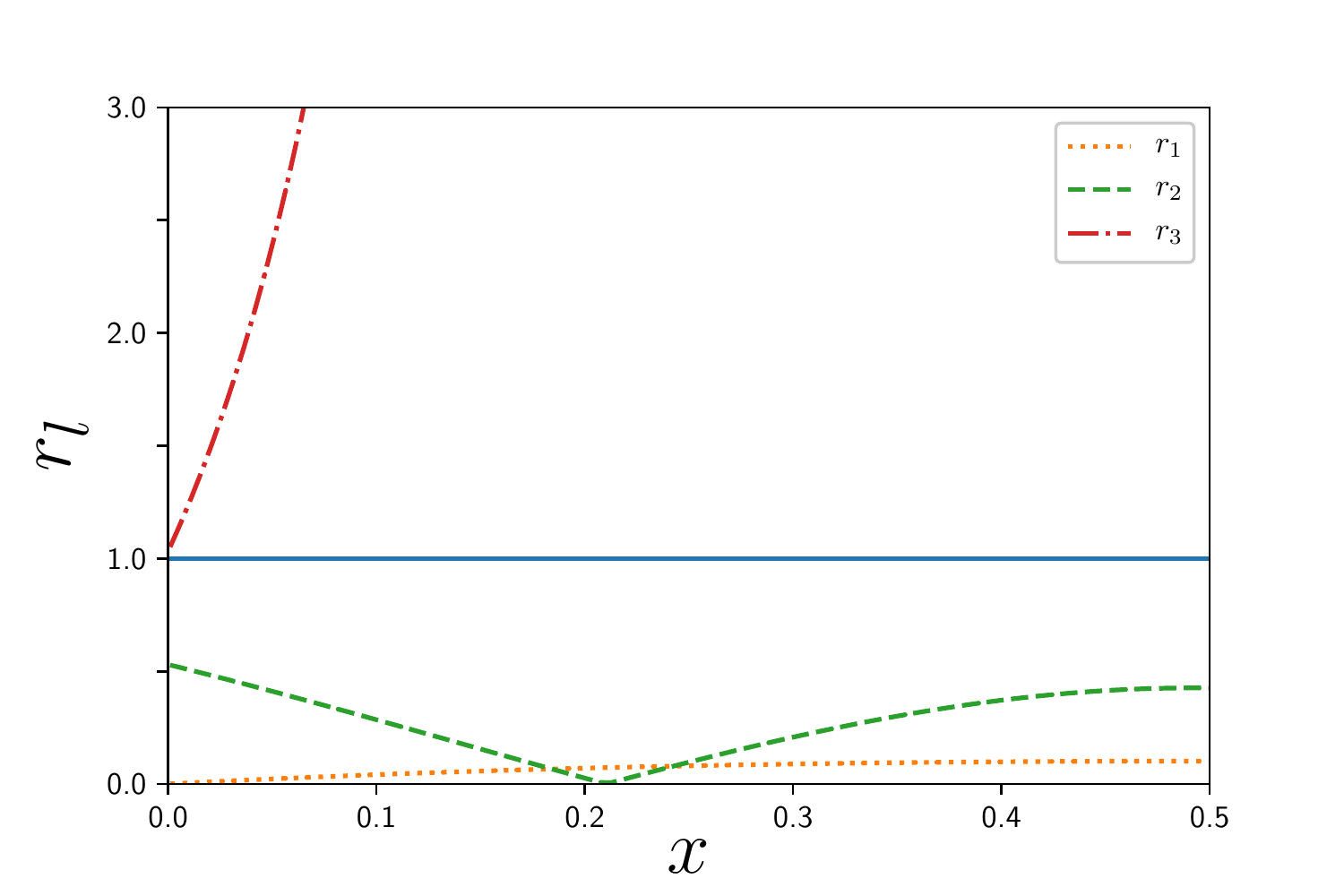}
\caption{Plot of the coefficient $r_l$ defined in eq. \eqref{eq:rl}. The horizontal bar at $r = 1$ is the maximum authorized value for $r_l$. If $r_l > 1$, then the corresponding term in the asymptotic series \eqref{eq:asymptotic_series} is increasing. }
\label{fig:rl}
\end{figure}

In this equation we have highlighted the fact that each step of the iteration brings out an additional factor of $b_0^{(l)} - b_0^{(l-1)}$ to $b_0$. Then the asymptotic series criterion requires us to cut the sum at the term at which $|b_0^{(l)} - b_0^{(l-1)}|$ is minimized in order to get the best precision on $b_0$. In Figure \ref{fig:rl} we have plotted the ratio
\begin{equation} \label{eq:rl}
r_l = \left| \frac{b_0^{(l)} - b_0^{(l-1)}}{b_0^{(l-1)} - b_0^{(l-2)}} \right| \; ,
\end{equation}
for $l=1, \; 2, \; 3$. The minimal term of the series is the one for which $r_l < 1$ and $r_{l+1} > 1$. We can see directly that, while the two first iterations of the procedure (i.e., matching up to the second derivative) seem to bring an improvement in the determination of $b_0$, the third one gives a large correction.

The end result of this Section is shown in Figure \ref{fig:b0}, where we plot the first four predictions $b_0^{(0)}, \; b_0^{(1)}, \; b_0^{(2)}$ and $b_0^{(3)}$ for $b_0$. Of course, an immediate drawback of this method is that we are not able to assess quantitatively the precision of the estimation of this coefficient ; moreover, it is apparent from Figure \ref{fig:b0} that even the qualitative behavior is not correct. A refinement of the method is needed but we leave this for future work.

\bibliography{Screened_potential.bib}

\end{document}